\documentclass[final]{elsarticle2}

\usepackage{lineno,hyperref}
\modulolinenumbers[5]

\newcommand{\partiel}[2]{\frac{\partial #1}{\partial #2}} 

\journal{European Journal of Mechanics B/Fluids}

\begin{document}

\begin{frontmatter}

\title{Optimal wavy surface to suppress vortex shedding \\ using second-order sensitivity to shape changes}
\date{6 December 2015}
\author{O. Tammisola}
\address[mymainaddress]{Dept.\ M3, The University of Nottingham,
University Park, Nottingham NG7 2RD UK}
\address[mysecondaryaddress]{KTH Mechanics, KTH Royal Institute of Technology, SE-10044 Stockholm,Sweden}

\begin{abstract}
A method to find optimal $2^{nd}$-order perturbations is presented, and applied to find the optimal spanwise-wavy surface for suppression of cylinder wake instability. Second-order perturbations are required to capture the stabilizing effect of spanwise waviness, which is ignored by standard adjoint-based sensitivity analyses. Here, previous methods are extended so that (i) $2^{nd}$-order sensitivity is formulated for base flow changes satisfying linearised Navier-Stokes, and (ii) the resulting method is applicable to a 2D global instability problem. This makes it possible to formulate $2^{nd}$-order sensitivity to shape modifications. Using this formulation, we find the optimal shape to suppress the a cylinder wake instability. The optimal shape is then perturbed by random distributions in full 3D stability analysis to confirm that it is a local optimal at the given amplitude and wavelength. Furthermore, it is shown that none of the 10 random wavy shapes alone stabilize the wake flow at $Re=50$, while the optimal shape does. At $Re=100$, surface waviness of maximum height 1$\%$ of the cylinder diameter is sufficient to stabilize the flow. The optimal surface creates streaks by passively extracting energy from the base flow derivatives and effectively altering the tangential velocity component at the wall, as opposed to spanwise-wavy suction which inputs energy to the normal velocity component at the wall. This paper presents a fully two-dimensional and computationally affordable method to find optimal $2^{nd}$-order perturbations of generic flow instability problems and any boundary control (such as boundary forcing, shape modulation or suction). 
\end{abstract}


\end{frontmatter}

%
%
\section{Introduction}
Since a decade, spanwise waviness is known to efficiently suppress vortex shedding and reduce drag behind bluff bodies. \cite{bearmanowen} showed experimentally that a spanwise wavy trailing edge completely suppressed the vortex shedding around a rectangular cylinder at $Re=40000$, resulting in a 30 $\%$ reduction of the mean drag. A similar effect was observed by \cite{darekarsherwin} numerically at $Re=100-500$. 

As pointed out by \cite{darekarsherwin}, the stabilizing effect of spanwise waviness may also be created by changing the wall boundary condition by bleed or transpiration. Through steady spanwise-alternating suction and blowing, \cite{kimchoi2005} shifted the Hopf bifurcation of the wake behind a circular cylinder from $Re\approx45$ to $Re>140$ in DNS. The instability could only be suppressed when the actuation had a spanwise wavelength of $5-6$ cylinder diameters. The reason for the efficiency of medium wavelengths has been analysed in several subsequent works. \cite{hwang} examined the instability of a fixed wake profile superposed with spanwise waviness, and observed that in this model medium wavelengths were not absolutely unstable. \cite{delguercio2014,delguercio2014b} considered base flow modifications generated by spanwise-alternating suction. They concluded that the streaks generated by suction were optimally amplified by transient growth at medium wavelengths, and hence the base flow modification was also largest at medium wavelengths. \cite{WavyWakes2014} considered formally modifications of global mode eigenvalues with spanwise-wavy base flow modifications, which required $2^{nd}$-order perturbations. Wavelength selection was based on an \textit{eigenmode resonance} at long wavelengths, and the strongest interaction with the $2^{nd}$-order sensitivity core at medium wavelengths.  

The optimal distribution of spanwise waviness has been studied much less than the optimal wavelength. However, for the flow around the circular cylinder, azimuthal location of the waviness is an important parameter. \cite{kimchoi2005} applied the spanwise-alternating suction from two slots placed on the top and the bottom of the cylinder; locations at the rear and front of the cylinder were mentioned to be inefficient. Moreover, the configuration in which the suction through the upper slot was in-phase with that through the lower slot was found to be much more effective than the anti-phase configuration, which was later explained using the mode resonance effect in \cite{WavyWakes2014}. \cite{delguercio2014b} performed a 3D optimization of the azimuthal distribution of waviness in order to create strongest possible base flow streaks. Their optimal distribution also peaked at the top and bottom of the cylinder but was continuous, and stabilized the flow at a much lower peak suction amplitude ($<1 \%$) than the slots of \cite{kimchoi2005} ($8 \%$) at $Re=100$. However, the optimization was performed on the streakiness of the base flow, and eigenvalue drift was not a part of the optimization. 

\cite{boujo2015} computed optimal spanwise-wavy base flow modifications for a parallel flow in a mixing layer, accounting for the eigenvalue drift. The $2^{nd}$-order perturbation system was written in matrix form and elegantly manipulated to form a Hessian matrix, and the most stabilizing perturbation found from its extremal eigenpairs. The manipulations involved forming an explicit inverse of a system matrix, which was possible since the flow was parallel with 1D eigenfunctions. The global wake instability problem considered here, however, has 2D eigenfunctions. 

The present study introduces a new approach to compute optimal boundary perturbations at the $2^{nd}$ order, accounting for both base flow change and eigenvalue drift. The perturbation system is projected on a smaller basis of boundary functions, and the optimal recovered using only 2D computations no larger than the original system. Using this method, we find the optimal spanwise-wavy cylinder surface to suppress vortex shedding around it.  Spanwise-wavy shapes are already used to suppress vortex shedding around \textit{e.g} chimneys. The optimal spanwise-wavy shape, however, has not been examined yet. 

The new attributes of this approach can be summarised as follows. In \cite{WavyWakes2014}, base flow modifications induced by wall suction were computed and analysed a posteriori. In \cite{boujo2015}, $2^{nd}$ order optimal base flow modifications were computed a priori. This base flow sensitivity was the $2^{nd}$-order counterpart of generic base flow sensitivity\cite{Marquet:2008p2300}, in particular, the base flow modifications did not satisfy Navier--Stokes equations. The present base flow modifications satisfy (linearised) Navier-Stokes equations, coupling this with the maximal eigenvalue drift, similarly to an adjoint base flow approach\cite{Marquet:2008p2300}. In addition, the projection to boundary basis functions makes it possible to apply the optimisation to two-dimensional problems. 
\section{Perturbation analysis \label{sec:pert}}
Let us consider a general eigensystem of the form:
\begin{equation}{\mathcal{L}_0 \{\textbf{q}_0\}=\sigma_0 \textbf{q}_0,}\end{equation}
where $\textbf{q}_0$ is an eigenvector, and $\sigma_0$ an eigenvalue. After introducing a small boundary modification denoted by $h$, we write:
\begin{equation}{\mathcal{L}(\epsilon h)\left\{\textbf{q}(h)\right\}=\sigma(h)\textbf{q}(h)\label{eq:hprob}}\end{equation}
In what follows, it will be assumed that the operator perturbation is linear in $h$: $\mathcal{L}(\epsilon h)=\mathcal{L}_0+\delta \mathcal{L}(h)$ . Physically this means that the base flow change is linear in $h$, which should be a reasonable assumption as long as $h$ is small (which is confirmed in Sec.\ \ref{sec:res}). If eigenvalue drifts are quadratic in $h$, this implies that base flow change is  linear in h; It has been shown previously that the eigenvalue drifts are quadratic in the base flow change \cite{WavyWakes2014,hwang}\footnote{If the assumption of linear base flow modifications is not valid, and the eigenvalue drift is not quadratic, then this just means that several iterations are needed  to find the optimal --- a familiar situation from gradient-based optimization.}.
The solution may be expanded in a perturbation series where $\epsilon$ denotes the amplitude of $h$ (\textit{e.g.} \cite{hinch}):
\begin{equation}{\left(\mathcal{L}+\delta\mathcal{L}\right) \left\{\sum_{n=0}^{2} \epsilon^n \textbf{q}_n+O(\epsilon^3)\right\}=\left(\sum_{n=0}^2 \sigma_n+O(\epsilon^3)\right) \left(\sum_{n=0}^2 \epsilon^n \textbf{q}_n+O(\epsilon^3)\right)\label{eq:pertexpprob}}\end{equation}
%
%
By grouping together terms of any given power of $\epsilon$, we can generate approximations of the eigenvalue drift accurate up to that order. At the first order in $\epsilon$:
\begin{equation}{\left(\mathcal{L}_0-\sigma_0 \mathcal{I} \right)\{\textbf{q}_1\}= -\delta \mathcal{L}\{\textbf{q}_0\}+\sigma_1 \textbf{q}_0, \label{eq:1ordrearr}}\end{equation}
where $\mathcal{I}$ is the identity operator. By projecting this equation under inner product $\langle,\rangle$ with the adjoint eigenmode $\textbf{q}_0^+$, the first order eigenvalue drift $\sigma_1$ is found to be:
\begin{equation}{\sigma_1= \langle \textbf{q}_0^+,\delta \mathcal{L} \{\textbf{q}_0\}\rangle, \label{eq:lambda1corr}}\end{equation}
which is equivalent to the \textit{sensitivity} used to estimate an eigenvalue drift with respect to control in numerous previous studies (see \textit{e.g.} \cite{annrevluchinibottaro} for a review). 
For spanwise wavy base flow modulations, $\sigma_1$ is known to equal zero \cite{hwang,delguercio2014}. Hence, the $2^{nd}$-order eigenvalue drift $\sigma_2$ is needed to estimate the value of control. 
In \cite{WavyWakes2014}, it was shown that $\sigma_2$ takes the form:
\begin{equation}{\sigma_2=\langle \textbf{q}_0^+,\delta \mathcal{L}\{\textbf{q}_1\}\rangle. \label{eq:lambda2corr}}\end{equation}
where $\textbf{q}_1$ is the first order eigenvector correction obtained from (\ref{eq:1ordrearr}).

\paragraph{\textbf{Optimal $2^{nd}$-order boundary perturbations}}
%
Shape changes, boundary suction, or mass injection at the cylinder can all be addressed by the method presented next with minimal adjustments to boundary conditions. 
It is common in shape optimization to parameterize the boundary, to reduce the degrees of freedom, but also obtain robust optimal shapes which are easy to manufacture \cite{jameson}. Let us parameterize the displacement of the cylinder wall using $N$ basis functions:
\begin{equation}{\delta h=\sum_{n=1}^{N} a_n h_n \label{eq:decomp}}\end{equation}
By substituting the above sum into (\ref{eq:lambda2corr}):
\begin{equation}{\sigma_2=\langle \textbf{q}_0^+,\delta \mathcal{L}\left(h\right)\left\{\textbf{q}_1(h)\right\}\rangle=\sum_{n=1}^{N} \sum_{m=1}^{N} a_n a_m   \langle \textbf{q}_0^+,\delta \mathcal{L}(h_n)\{\textbf{q}_1(h_m)\}\rangle=\sum_{n=1}^{N} \sum_{m=1}^{N} a_n a_m \hat S_{nm} \label{eq:lambda2corrsum}}\end{equation}
The sum could be moved outside the inner product since $\delta \mathcal{L}(h)$ is linear in $h$ (and since Eq.\ \ref{eq:1ordrearr} is linear in $\delta \mathcal{L}$, $\textbf{q}_1(h)$ is also linear in $h$). 
The complex matrix $\hat \textbf{S}$ is the projected form of the sensitivity operator \citep{boujo2015} acting on the boundary basis coefficients, and hence can directly reproduce the eigenvalue drift for any chosen combination of boundary basis functions. The computation of the optimal drift, however, remains to be addressed. If $\hat \textbf{S}$ were a symmetric matrix, this computation would be simple. The eigenvectors of complex symmetric matrices are orthogonal (a known result in linear algebra, see \textit{e.g.} \cite{heath}). This means that the largest eigenvalue drift $\sigma_2=\textbf{a}^T \tilde S \textbf{a}$ is achieved when the coefficient vector $\textbf{a}$ is an eigenvector of $\tilde \textbf{S}$, and more precisely, the eigenvector of the largest magnitude eigenvalue of $\tilde \textbf{S}$. Similarly, the largest decrease in growth rate (most negative $\sigma_{2,r}$) is achieved for the eigenvalue of $\tilde \textbf{S}$ with the most negative real part. 

Next, we show that the action of $\hat \textbf{S}$ on $\sigma_2$ can be rewritten using a symmetric sensitivity operator $\tilde\textbf{S}$.  To see this, set $\tilde{S}=0.5(\hat S+\hat S^T)$ to obtain $\sigma_2=\sum_{n=1}^{N} \sum_{m=1}^{N}a_n \hat S_{nm} a_m=\sum_{n=1}^{N} \sum_{m=1}^{N}a_n a_m \tilde{S}_{nm}$. Moreover, the element $\tilde{S}_{nm}$ is simply the $2^{nd}$-order eigenvalue drift obtained when $a_m=a_n=1$:
\begin{equation}{\tilde{S}_{mn}= \sigma_2 \left(a_m=a_n=1, \quad a_j=0 \quad \forall j \neq m,n  \right) \label{eq:S}}\end{equation}
Summarizing, the elements of the sensitivity operator (matrix) $\tilde\textbf{S}$ are formed by computing eigenvalue drifts for $N \times N/2$ pairs of basis functions ($h_n,h_m$). The most stabilizing shape change at the $2^{nd}$ order is the eigenvector of the small matrix real($\tilde\textbf{S}$) corresponding to the most negative eigenvalue. The most destabilizing shape change at the $2^{nd}$ order is the eigenvector of the small matrix real($\tilde\textbf{S}$) corresponding to the most positive eigenvalue. 
%
%

It is worth noting that this method is fundamentally equivalent to the one in \cite{boujo2015}, where the sensitivity operator $\sigma_2= \langle \delta \textbf{U},\textbf{S}\{\delta \textbf{U}\} \rangle$ was formed. By utilizing the boundary parameterization and computing eigenvalue drifts, we avoid forming a matrix inverse of the operator $\mathcal{L}-\sigma_0 \mathcal{I}$, not possible in a 2D global stability problem.   

\begin{figure}
\vspace{-2.24cm}
\centering
\includegraphics{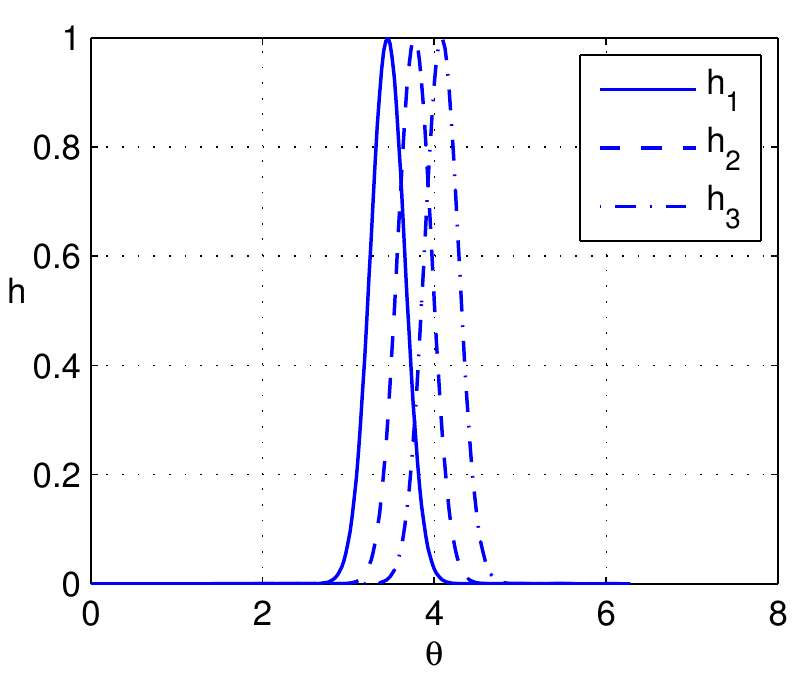} 
\caption{The three first surface modification basis functions: $h_1$, $h_2$ and $h_3$ (see legend). The modification happens in the surface-normal direction, \textit{i.e.} the radial direction ($\delta r (\theta)=h(\theta)$).}
\label{fig:hbasis}
\end{figure}
\begin{figure}
\centering
\includegraphics{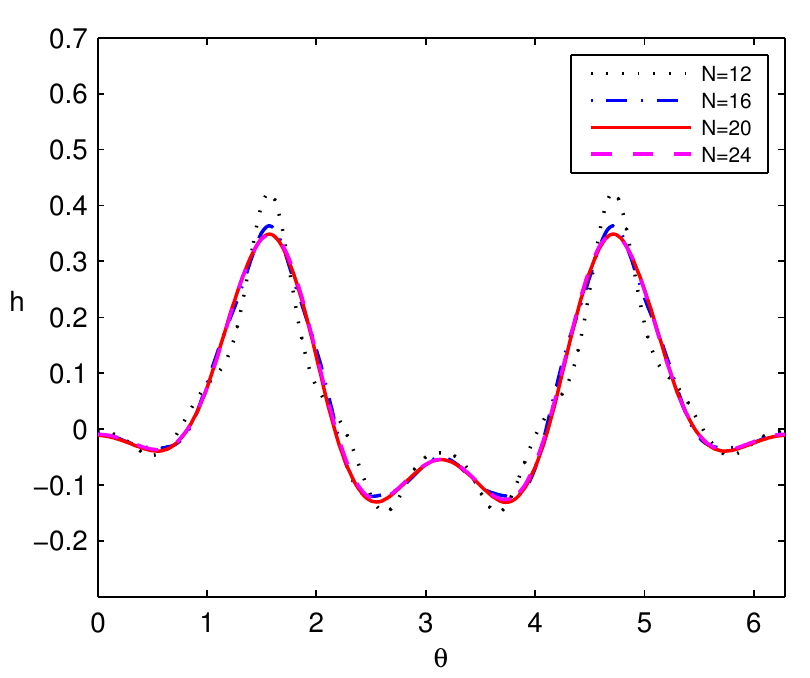} 
\caption{Convergence of the optimal shape at $Re=50$ with respect to the number of basis functions $N$.}
\label{fig:Nconv}
\end{figure}

\paragraph{\textbf{Application to a spanwise-wavy cylinder surface}}
This study considers the stabilization of the primary wake instability eigenmode by means of spanwise-wavy surface modifications. The eigenmodes $\textbf{q}_0=(\textbf{u}_0,p_0)$ for the flow around a straight circular cylinder satisfy the linearized, incompressible Navier--Stokes equations: 
\begin{equation}{\mathcal{L}_0\textbf{q}_0=\textbf{U} \cdot \nabla \textbf{u}_0+\textbf{u}_0 \cdot \nabla \textbf{U}+\nabla{p}_0-Re^{-1}\nabla^2 \textbf{u}_0=-\sigma_0 \textbf{u}_0,\label{eq:dir}}\end{equation}
where $\textbf{U}$ is the 2D non-wavy base flow. The adjoint eigenmodes $\textbf{q}^+_0=(\textbf{u}^+_0,p_0)$ satisfy 
\begin{equation}{\left(\nabla \textbf{U}\right)^T \cdot \hat \textbf{u}^+_0 - \textbf{U} \cdot \nabla \hat \textbf{u}^+_0 - \nabla \hat{p}^+_0 - \frac{1}{Re} \nabla^2 \hat \textbf{u}^+_0=\sigma_0^*\textbf{u}_0,\label{eq:adj}}\end{equation}
where $^*$ is a complex conjugate. 

The wavy cylinder problem is described using of cylindrical coordinates in the explanation below, for simplicity. The final equations apply to normal displacement of arbitrary surface and were implemented in the code using Cartesian coordinates. The characteristic length scale of the problem is the non-wavy dimensional cylinder radius $r^{'}=a$, where $^{'}$ denotes a dimensional quantity. Therefore the nondimensional wavy cylinder radius becomes $r=(r^{'}/a)=1+h(\theta)$, where $h(\theta,z)$ is a wavy displacement in the surface-normal (radial) direction. 
We can express the modification of a point at the cylinder surface in a chosen surface basis function set as:
\begin{equation}{h(\theta,z)=A \sum_{n=1}^{N} a_n h_n (\theta, z)\cos\{\beta_B z\} \label{eq:dec}}\end{equation}

Introducing the boundary modification (\ref{eq:dec}) into the $2^{nd}$-order eigenvalue drift obtained in \cite{WavyWakes2014}, we find that \\$\sigma_2=\int_V -\hat \textbf{u}_0^{+*}\left(\hat \textbf{u}_{(1,+)} \cdot \nabla \delta \textbf{U}_- +\delta \textbf{U}_- \cdot \nabla \hat  \textbf{u}_{(1,+)}+\hat \textbf{u}_{(1,-)} \cdot \nabla \delta \textbf{U}_+ +\delta \textbf{U}_+ \cdot \nabla \hat \textbf{u}_{(1,-)}\right) dV
= \sum_{n=1}^{N} \sum_{m=1}^{N} a_n\tilde S_{nm}a_m,$ where
\begin{eqnarray}
\tilde{S}_{nm}& =&\int_V -\hat \textbf{u}_0^{+*}(\hat \textbf{u}_{(1,+)}(h_n) \cdot \nabla \delta \textbf{U}_-(h_m) +\delta \textbf{U}_-(h_m) \cdot \nabla \hat  \textbf{u}_{(1,+)}(h_n) \nonumber \\
&& +\hat \textbf{u}_{(1,-)}(h_n) \cdot \nabla \delta \textbf{U}_+(h_m) +\delta \textbf{U}_+(h_m) \cdot \nabla \hat \textbf{u}_{(1,-)}(h_n)) dV  \nonumber \\
&& \int_V -\hat \textbf{u}_0^{+*}(\hat \textbf{u}_{(1,+)}(h_m) \cdot \nabla \delta \textbf{U}_-(h_n) +\delta \textbf{U}_-(h_n) \cdot \nabla \hat  \textbf{u}_{(1,+)}(h_m) \nonumber \\
&& +\hat \textbf{u}_{(1,-)}(h_m) \cdot \nabla \delta \textbf{U}_+(h_n) +\delta \textbf{U}_+(h_n) \cdot \nabla \hat \textbf{u}_{(1,-)}(h_m)) dV 
\label{eq:sigma2numerich}
\end{eqnarray}
In the above, $\delta \textbf{U}(h_n)=\delta \textbf{U}_+(h_n)\exp(\textrm{i}\beta_B z)+\delta \textbf{U}_-(h_n)\exp(-\textrm{i}\beta_B z)$ is the base flow modification obtained when the boundary function amplitudes are set to $h_n=1$ and $h_j=0 \quad \forall j \neq n$. Similarly, the eigenvector correction is a sum of two parts: $\textbf{u}_{1}(h_n)=\textbf{u}_{(1,+)}\exp(\textrm{i}\beta_B z)+\textbf{u}_{(1,-)}\exp(-\textrm{i}\beta_B z)$ . Here, $\textbf{u}_{(1,\pm)}(h_n)$ is the eigenvector correction (\ref{eq:1ordrearr}) obtained using the base flow modification $\delta \textbf{U}_{\pm}(h_n)$:
 \begin{eqnarray}
-\hat \textbf{u}_{\left(1,\pm \right)}\cdot \nabla \textbf{U} -\textbf{U} \cdot \nabla \hat \textbf{u}_{(1,\pm)}- \nabla \hat p_{1,\pm}+ Re^{-1}\nabla^2\hat \textbf{u}_{(1,\pm)}-\sigma\hat \textbf{u}_{(1,\pm)}&=& \nonumber \\
 \hat \textbf{u}_{(1,\pm)} \cdot \nabla \delta \textbf{U}_{\pm} +\delta \textbf{U}_{\pm}\cdot \nabla \hat \textbf{u}_{(1,\pm) \label{eq:evcorr}} %
\end{eqnarray}
The right-hand side in Eq.\ (\ref{eq:evcorr}) is yet unknown. To determine it, we need to know $\delta \textbf{U}_\pm(h_n)$ --- the relation between the shape modifications and the base flow modifications induced by them. This could be obtained by solving the steady Navier-Stokes equations for the wavy cylinder flow, and forming the difference between the base flows of the wavy cylinder and the straight cylinder, similarly to \cite{WavyWakes2014}. Such an approach presents a few issues for the wavy cylinders. A minor difficulty is interpolation between different flow domains. Most importantly, the method of $2^{nd}$ order base flow sensitivity in \cite{boujo2015} cannot be extended to $2^{nd}$ order shape sensitivity, if the relation $\delta \textbf{U}_\pm(h_n)$ is nonlinear. We are hence looking for a linear relation. This can be obtained from the \textit{linearised} steady Navier-Stokes equations with appropriate \textit{linearised} boundary conditions.
To obtain the correct linearised boundary condition, we introduce a Taylor-decomposition of the no slip condition for the waviness-modified base flow $\textbf{U}_{tot}=\textbf{U}+\delta \textbf{U}_\pm(h_n)$, at the wavy cylinder surface:
\begin{equation}{0=\textbf{U}_{tot}(r=1+h_n)=\textbf{U}_{tot}(r=1)+h_n\frac{\partial{\textbf{U}_{tot}}}{\partial r}(r=1)+O(h_n^2)}\end{equation}
Since $h_n\left(\delta \textbf{U}_\pm / \partial r\right)=O(h_n^2)$, after ignoring nonlinear terms we obtain:
\begin{equation}{0=\textbf{U}_{tot}(r=1)+h_n\frac{\partial{\textbf{U}}}{\partial r}(r=1)}\end{equation}
To obtain a boundary condition for $\delta \textbf{U}_\pm(h_n)$, we subtract from this the base flow condition $\textbf{U}(r=1)=0$ to get:
\begin{equation}{\delta \textbf{U}_\pm(h_n)=-h_n\frac{\partial{\textbf{U}}}{\partial r}(r=1)=-h_n (\nabla \textbf{U} \cdot \hat n)\label{eq:linbfmodBCprel}}\end{equation}
Concluding, the base flow modifications for each basis function, $\delta \textbf{U}_\pm (h_n)$ are found from the Navier-Stokes equation linearised around a straight cylinder base flow $\textbf{U}$ with respect to $h_n$ as:
\begin{equation}{\delta \textbf{U}_\pm(h_n)\cdot \nabla \textbf{U}+ \textbf{U} \cdot \nabla \delta \textbf{U}_\pm(h_n) =-\nabla \delta P_\pm(h_n)+Re^{-1}\nabla^2 \delta \textbf{U}_\pm(h_n),\label{eq:bfmod}}\end{equation} 
with the linearized boundary condition at the cylinder surface:  
\begin{equation}{\delta \textbf{U}_\pm(h_n)=-h_n (\nabla \textbf{U} \cdot \hat n)\label{eq:bfmodBC}}\end{equation} 
where $\hat n$ is the normal pointing into the cylinder, and $\delta \textbf{U}_\pm(h_n)=0$ at all other boundaries. 
Here, the gradients are given by: 
\begin{displaymath}{\nabla \textbf{U}_\pm(h_n)=\left[\partiel{\delta \textbf{U}_\pm(h_n)}{x},\partiel{\delta \textbf{U}_\pm(h_n)}{y},\pm \textrm{i} \beta_B \delta \textbf{U}_\pm(h_n)\right]}\end{displaymath}
\begin{displaymath}{\nabla \textbf{U}=\left[\partiel{\textbf{U}}{x},\partiel{\textbf{U}}{y},0\right]}\end{displaymath}

The above equations apply to arbitrary surface shapes. Here, we solve for the action of the gradient, $(\delta \textbf{U}/\delta h)\delta\textbf{U}$, directly without forming an adjoint, which is termed sensitivity-based method in \cite{gunzburger}.
\paragraph{Justification of Eq.\ (\ref{eq:bfmod}--\ref{eq:bfmodBC}) \label{par:just}} It is worth mentioning the additional reasons why linear base flow modifications have been adopted, even though the main reason is that a linear condition allows a generalisation of the method of \cite{boujo2015} to base flow modifications induced by boundary perturbations. 

Firstly, the same boundary condition is common in shape receptivity studies of boundary layers (\textit{e.g.} \cite{bertolotti,Tempelmann}); in cases where the receptivity remains linear, then also the base flow modification resulting from this boundary condition is linear. Likewise, the linearisation of velocity with surface height is routinely performed in linear stability studies of free-surface flows (\textit{e.g.} \cite{sasakipeket,StableWe,VakarWe}). Also, an assumption of linear base flow modifications is always made in $1^{st}$ order sensitivity studies with respect to volume forcing (\textit{e.g.} \cite{Marquet:2008p2300}, \cite{PRALITS:2006p2515}), and in sensitivity to boundary forcing (\textit{e.g.} \cite{Xjunction}). 

In (\ref{eq:bfmodBC}), a velocity boundary condition is applied at the unperturbed surface. This might lead one to think that optimal wavy shape is the same as optimal wavy wall-normal suction, or that the two can be obtained from each other. Such equivalence is proved wrong by previous receptivity studies, as well as the present work. Boundary-layer receptivity to shape changes is different from receptivity to wall suction, even when linearised velocity boundary conditions are applied at the surface; this is shown in \textit{e.g.} \cite{bertolotti}, where both cases are analysed separately. The reason for this is quite simple. Suction velocity at the boundary is wall-normal, while both the direction and the magnitude of the velocity induced by shape changes at the boundary depends on the local base flow gradients, as Fig.\ \ref{fig:indvel} in Sec. \ref{sec:res} illustrates. 

Secondly, from a theoretical point of view, it is worth noting that the assumption of linear base flow changes with height is actually not an additional assumption, but is embedded into the main assumption that eigenvalue drift induced by a boundary perturbation is of the $2^{nd}$ order. It is known that the eigenvalue drift induced by base flow modifications is $2^{nd}$ order with respect to the base flow modification amplitude \cite{hwang,WavyWakes2014,boujo2015}. Because of this, only a linear relation $\textbf{U}_\pm(h_n)$ will enforce an eigenvalue drift of the $2^{nd}$ order. Any nonlinear contributions in the $\textbf{U}_\pm(h_n)$ relation will only influence the eigenvalue drift at the $3^{rd}$ and higher orders. Hence, if the eigenvalue drift is of the $2^{nd}$ order with respect to a boundary perturbation, then the assumption of linear base flow changes must be implicitly satisfied. In any case, the analysis presented here aims to find the wavy surface generating maximal $2^{nd}$-order eigenvalue drift, and a linear base flow modification is consistent with this aim. 
\section{Numerical method \label{sec:num}}
To formulate the problem numerically, we start by choosing $N$ basis functions onto which the surface modification can be projected. The method in this paper can be perfomed with an arbitrary set of basis functions. The ideal basis is problem-dependent, the choice between local and global basis functions is determined by a tradeoff between flexibility and accuracy. In the present work, a local basis of Gaussians has been chosen to describe the normal (radial) displacement of the surface point $\textbf{r}=\left[cos{\theta}, sin{\theta}\right]$ (where $\theta$ is an azimuthal angle) as:
\begin{equation}{h_n=\exp\{-50 ||\textbf{r}-\textbf{r}_{0,n}||\}\cos\{\beta_B z\}.}\end{equation}
The midpoints of the Gaussians are evenly spaced over the cylinder circumference:
 \begin{equation}{\textbf{r}_{0,n}=\left[\cos(2\pi n/N),\sin(2\pi n/N)\right]}\end{equation} 
As shown in Fig.\ \ref{fig:hbasis}, the chosen basis imitates a finite element basis, as each function has in practice a finite support. Convergence of the optimal shape at $Re=50$ with respect to the number of basis functions is shown in Fig.\ \ref{fig:Nconv}. The shapes with $N\ge 20$ and $N=24$ are very close, and hence $N=20$ was considered to be sufficient. The Fourier basis used to discretise boundary suction distributions in \cite{delguercio2014b} might require a smaller number of basis functions for a converged optimal, if the optimal solution does not vary rapidly.

The numerical solution of the optimal wavy shape is fully two-dimensional and consists of five steps: (a) base flow, (b) direct and adjoint eigenmodes (\ref{eq:dir} \& \ref{eq:adj}), (c) base flow modifications ($N$ solutions of \ref{eq:bfmod}), (d) eigenvector corrections ($N$ solutions of \ref{eq:evcorr}), (e) inner products (evaluations of the 0.5$N^2+0.5N$ integrals in \ref{eq:lambda2corrsum}).  
%
%
For simplicity, the whole method is implemented into the matrix-based open-source finite-element software FreeFem++ (a more detailed description of the method can be found in \cite{WavyWakes2014}). It is worth mentioning that the method is software-independent and could be implemented in connection to any linear stability solver (whether matrix-based or a time-stepper such as Nek5000).

After derivation of the variational formulation of the governing equations, the associated sparse matrices are built by FreeFem++. The grid consists of 97010 Taylor-Hood elements ($\textbf{P}_2-\textbf{P}_1$). 
Steps (a) and (b) are identical to \cite{WavyWakes2014}. In (a), the base flow for the non-wavy cylinder is generated by solving the time-independent (steady) Navier-Stokes equations using a Newton-Raphson method. In (b), the direct and adjoint eigenproblems for the non-wavy cylinder are solved by the Arnoldi solver in FreeFem++ combined and UMFPACK. In step (c), the steady linear equation systems (\ref{eq:evcorr}) for the base flow modifications are solved directly using the sparse LU solver UMFPACK.  In step (d), the steady eigenvector correction equations (\ref{eq:evcorr}) are solved again with UMFPACK. These computations are needed to form the small $N \times N$-matrix $\tilde\textbf{S}$. The eigenpairs of this small matrix can be found using any standard QZ-algorithm, \textit{e.g.} \textbf{eig} of MATLAB used in the present work. The reason a QZ-algorithm is preferred over other methods is that $\tilde\textbf{S}$ may be nearly singular in the general case. The eigenvalues of $\tilde\textbf{S}$ represent global mode eigenvalue drifts for different basis function combinations, and many combinations may have negligible effect on the eigenvalue (resulting in near-zero eigenvalues of $\tilde\textbf{S}$). 

The verification of the above method is performed by a direct computation of 3D base flows and tri-global eigenmodes with a wavy boundary using a spectral element method (SEM) implemented in Nek5000 \cite{nek5000code}, which has an efficient parallelization scaling linearly up to millions of processors. The grids for the wavy cylinders are generated by starting from a mesh for a flow around a straight cylinder. The outer domain boundaries form a rectangular block with a diagonal from $[x,y,z]=[-20,-20,-\pi/\beta_B]$ to $[x,y,z]=[50,20,\pi/\beta_B]$. The element distribution is uniform in $z$, and finest close to the cylinder in the $x$-$y$-plane. Periodic boundary conditions are imposed at $z=\pm\pi/\beta_B$. In this work, two different meshes have been constructed: one for $\beta_B=0.8$ (52240 spectral elements, 12 elements in $z$) and $\beta_B=1.26$ (34817 elements, 8 elements in $z$). Another option would have been to stretch the same mesh in the $z$-direction. In a second step, the mesh is mapped onto the wavy cylinder boundary by solving a Laplace equation for the mesh displacements in the $x$-$y$-plane, where the boundary conditions are a given displacement at the cylinder boundary and zero displacement at the outer boundaries. 
Firstly, the base flows around wavy cylinders are obtained by integrating the nonlinear Navier-Stokes equation forward in time, and converging towards the steady solution by applying selective frequency damping \cite{sfd}. 
Secondly, the full 3D eigenpairs of the Linearized Navier--Stokes operator are computed using 
the linearized DNS time-stepper available in Nek5000 coupled with an Arnoldi method as in \cite{WavyWakes2014}. 
\begin{figure}
\hspace{-0.15cm}
\includegraphics{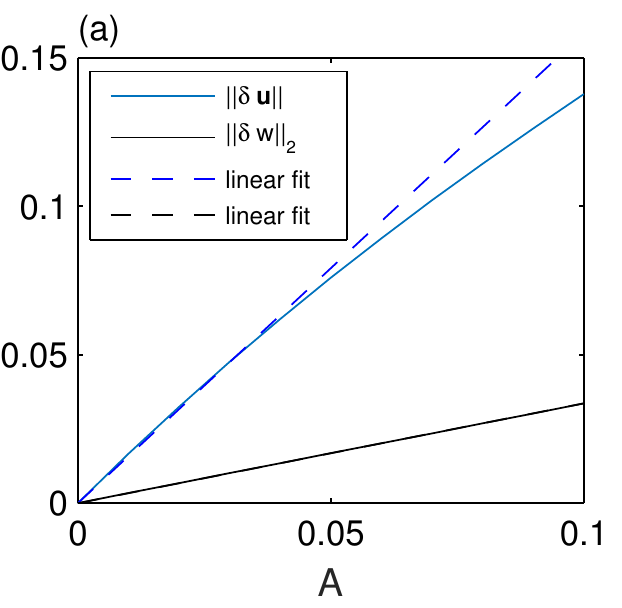}
\hspace{0.2cm}
\includegraphics{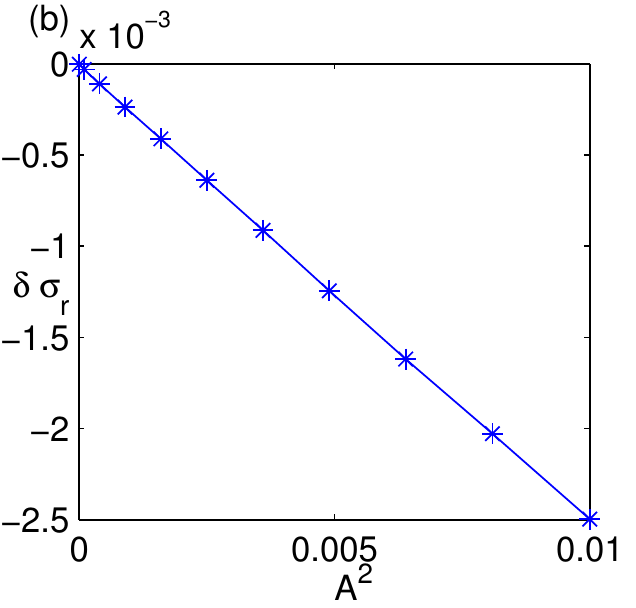} \\ 
\hspace{-0.3cm}
\includegraphics{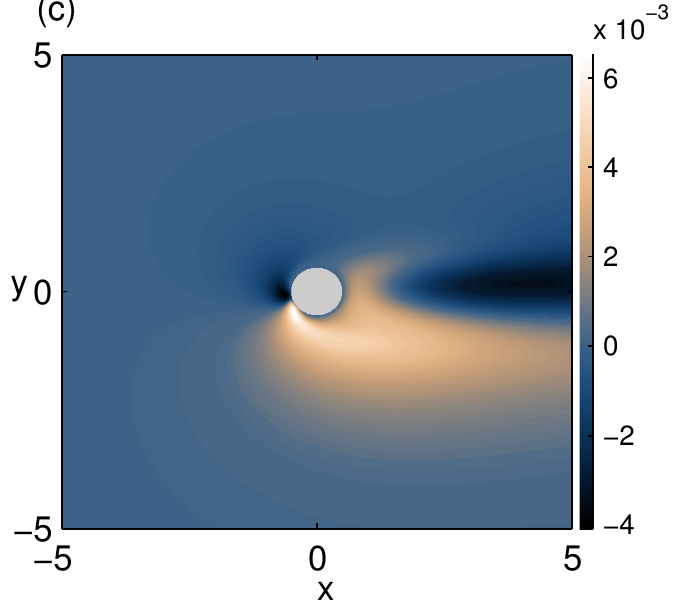}
\hspace{-0.2cm}
\includegraphics{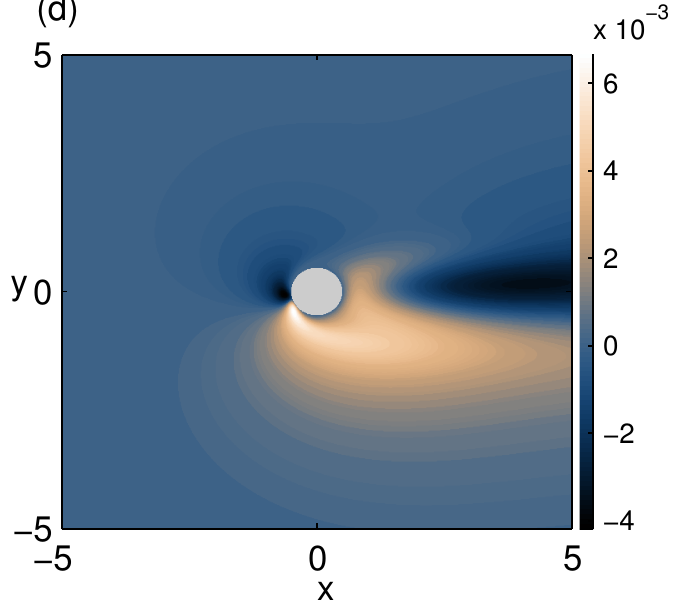}
\caption{Validation of the method using basis function $h_1$ ($a_1=A$ and $a_j=0 \forall j \neq 1$): (a) $L_2$-norm of the base flow change. This indicates that the base flow change is approximately linear up to amplitudes $A\approx0.1$. (b) The eigenvalue drift with respect to $A^2$. This indicates that the eigenvalue drift is quadratic up to amplitudes $A\approx0.1$. (c) Exact difference between base flows with and without surface waviness, spanwise velocity. (d) Predicted difference (FreeFem++).}
\label{fig:Nekquadratic}
\vspace{-0.2cm}
\end{figure}
\section{Results \label{sec:res}}
The basic method for the base flow and eigenmode computations has been cross-validated between the two solvers (FreeFem++ and Nek5000), and against results from the literature. For example at $Re=50$, $\sigma=0.014+0.752\mathrm{i}$ ($St=0.200$) for FreeFem++\footnote{This value contained a typo in the previous version}, and $\sigma=0.0128+0.746 \mathrm{i}$ ($St=0.119$) for Nek5000. Both compare very well to the eigenvalue in Fig.\ 7 at $Re=50$ in \cite{giannetti} with $\sigma_r=0.013$ and $St=0.119$. The method of $2^{nd}$-order perturbations was introduced in \cite{WavyWakes2014}, and the eigenvalue drift reproduced well against 3D computations. The present work introduces two new features to the method: (1) Optimal $2^{nd}$-order perturbations, and (2) computation of base flow modifications inside the algorithm. The optimality problem can hence be solved entirely in 2D, without computing an explicit matrix inverse as in \cite{boujo2015}, and without relying on 3D computations of the base flow change as in \cite{WavyWakes2014} or \cite{delguercio2014b}.
Here, we focus on validating these new features. The method assumes that base flow modification is linear with changing amplitude of $h$ (\ref{eq:bfmod}), which is valid when $h$ is sufficiently small. It would be impossible to test this assumption for all combinations of the basis functions at all amplitudes and wavelengths; each validation case in 3D (a base flow and eigenmode computation at a given waviness distribution and amplitude) required between 6000-10000 CPU hours. Instead, we focus on basis function $h_1$ at $\beta_B=0.8$, and vary its amplitude. Firstly, a 3D base flow around the wavy cylinder is computed in Nek5000. The difference is then formed between the base flow velocity around the wavy cylinder, and the base flow velocity around the non-wavy cylinder. This gives the exact base flow modification. Fig.\ \ref{fig:Nekquadratic} (a) shows the $L_2$-norm of the base flow velocity modification as function of the amplitude of the waviness $A=||\textbf{a}||$, together with a linear fit, up to as high value as $A=0.1$, a surface displacement corresponding to 10$\%$ of the cylinder diameter. The norm is nearly linear with amplitude up to around $A=0.05$, and curves away slightly at $A>0.05$. This shows that the base flow modifications are essentially linear for all amplitudes investigated in this paper. The $w$-component (also shown) is linear up to $A=0.1$. Furthermore, to validate the numerical solution, the shape of the exact base flow modifications from Nek5000 (\ref{fig:Nekquadratic} c) is compared against the shape of the linear base flow modifications from FreeFem++ (\ref{fig:Nekquadratic} d). We don't expect an exact match due to two different numerical methods; neverthless, the distribution and amplitude (colorbars) are very similar. As discussed in Sec.\ \ref{sec:pert}, the assumption of linear base flow modifications is also embedded into the assumption of quadratic eigenvalue drifts. 
The 3D eigenvalues computed in Nek5000 (Fig.\ \ref{fig:Nekquadratic} b) form a line against $A^2$ showing that the eigenvalue drifts are quadratic, and thus further confirms that our basic assumptions are valid. 
\begin{table}{
\centering
{\begin{tabular}{ c|c|c|c}
 Case  & $ \sigma$  \\
\hline
     Original straight cylinder & $0.0128+0.7467\mathbf{i}$\\
     Optimal unperturbed & $-0.0235+0.7314\mathbf{i}$ \\
     Optimal perturbed by random 1 & $-0.0206+0.7310\mathbf{i}$  \\
     Optimal perturbed by random 2 & $-0.0205+0.7313\mathbf{i}$  \\
     Optimal perturbed by random 3 & $-0.0211+0.7308\mathbf{i}$  \\
     Optimal perturbed by random 4 & $-0.0209+0.7315\mathbf{i}$   \\
     Optimal perturbed by random 5 & $-0.0204+0.7314\mathbf{i}$   \\
     Optimal perturbed by random 6 & $-0.0208+0.7355\mathbf{i}$   \\
     Optimal perturbed by random 7 & $-0.0206+0.7357\mathbf{i}$   \\
     Optimal perturbed by random 8 & $-0.0209+0.7351\mathbf{i}$   \\
     Optimal perturbed by random 9 & $-0.0223+0.7307\mathbf{i}$   \\
     Optimal perturbed by random 10 & $-0.0211+0.7308\mathbf{i}$   \\
     Random 1 & $0.0130+0.7558\mathbf{i}$    \\
     Random 2 & $0.0129+0.7462\mathbf{i}$    \\
     Random 3 & $0.0098+0.7433\mathbf{i}$    \\
     Random 4 & $0.0058+0.7438\mathbf{i}$    \\
     Random 5 & $0.0101+0.7452\mathbf{i}$    \\
     Random 6 & $0.0254+0.7462\mathbf{i}$   \\
     Random 7 & $0.0131+0.7462\mathbf{i}$  \\
     Random 8 & $0.0129+0.7463\mathbf{i}$   \\
     Random 9 & $0.0007+0.7412\mathbf{i}$   \\
     Random 10 & $0.019+0.7455\mathbf{i}$  \\
     \end{tabular}}
\caption{Eigenvalues from 3D computations in Nek5000 of: (a) the optimal wavy shape normalized to $A=0.04$ (row 1), (b) the optimal wavy shape superposed with a random perturbation at $A=0.01$ and re-normalized to $A=0.04$ (rows 2-11), and (c) the random shapes normalized to $A=0.04$ (rows 12-21). The 20 coefficients of the random vectors are obtained in succession from MATLAB:s \textbf{rand}-function with default settings.}
\label{table:t1}
}\end{table}

%
%
\begin{figure}
\centering
\hspace{1cm}
\includegraphics[width=0.25\textwidth]{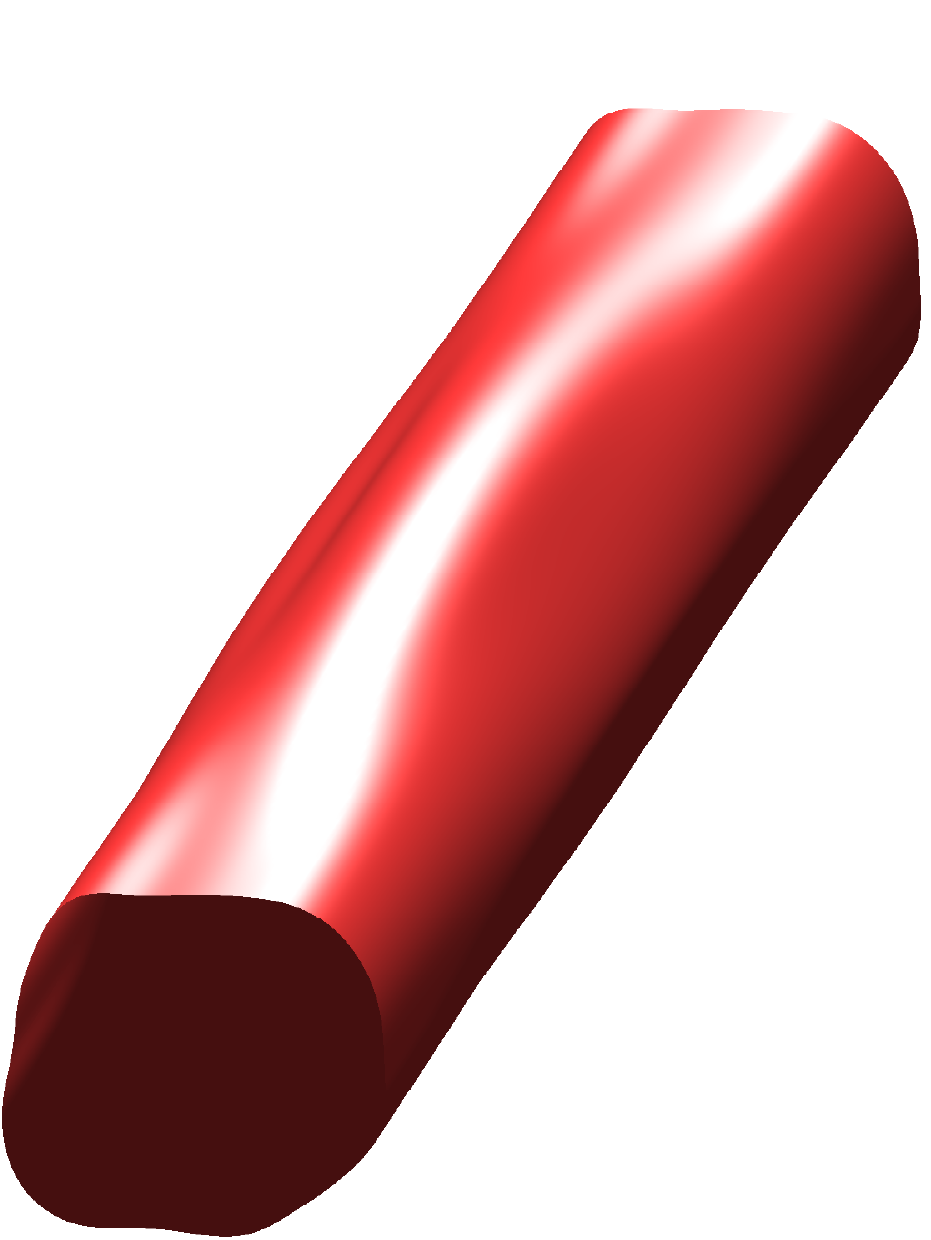}
\includegraphics[width=0.25\textwidth]{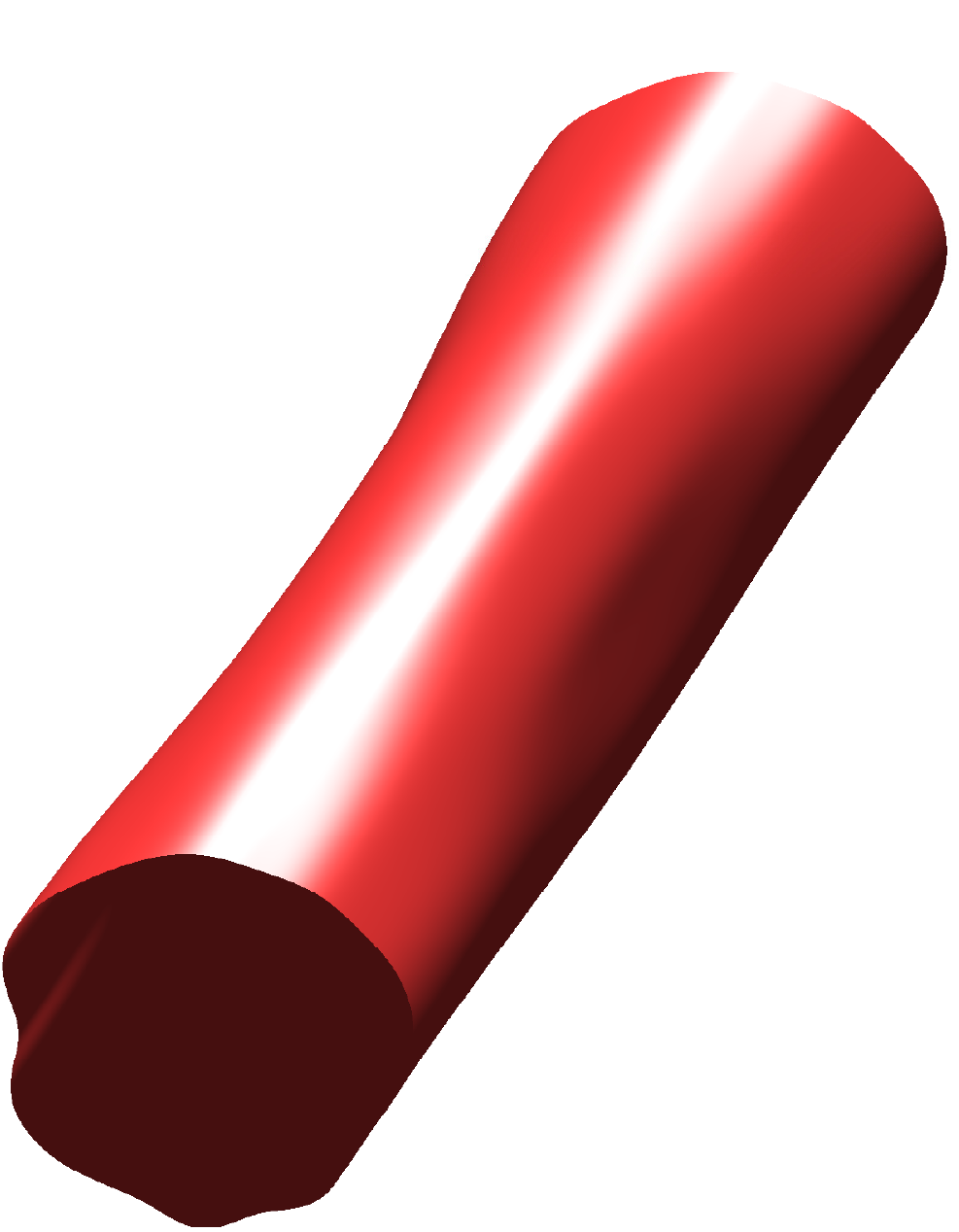}
\includegraphics[width=0.25\textwidth]{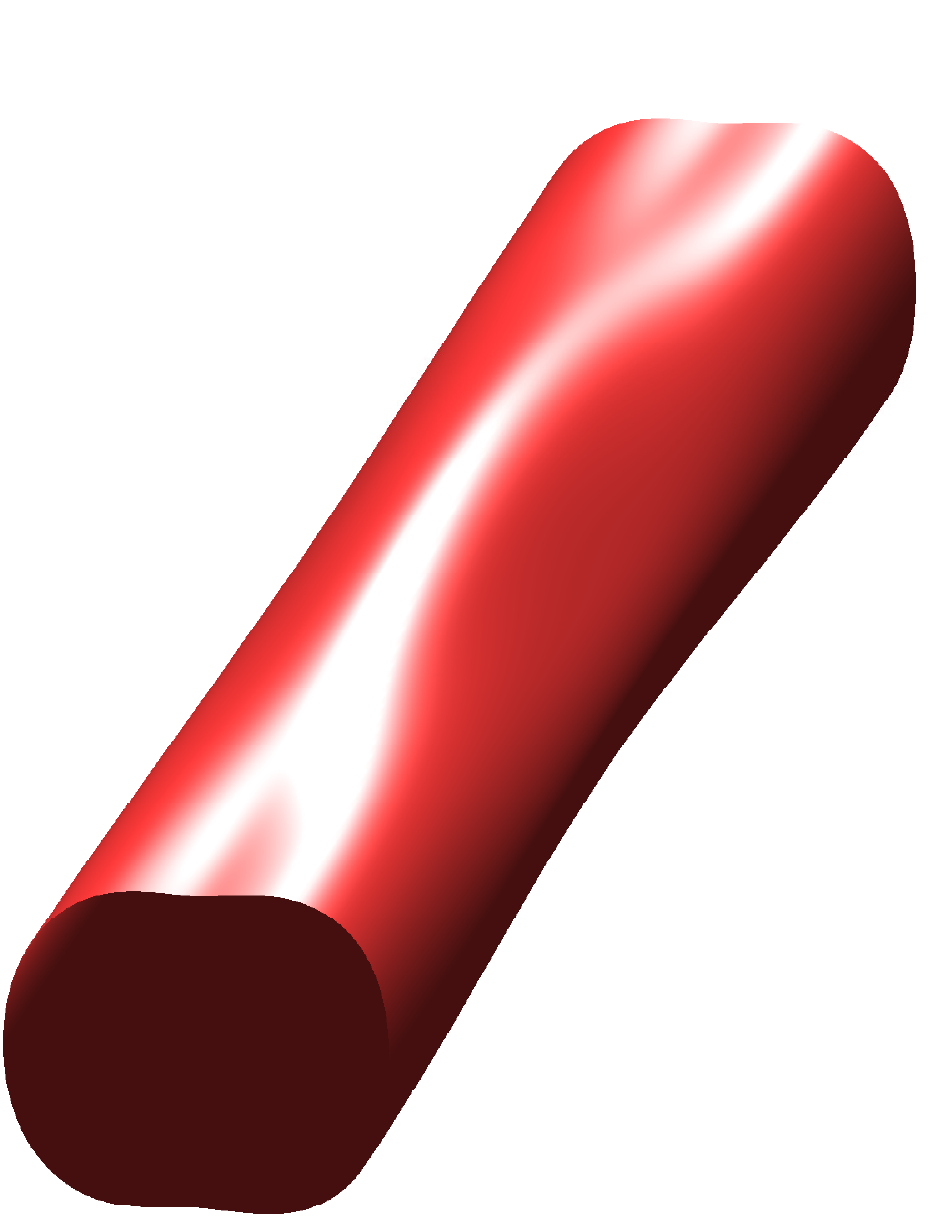} \\ 
\includegraphics[width=0.25\textwidth]{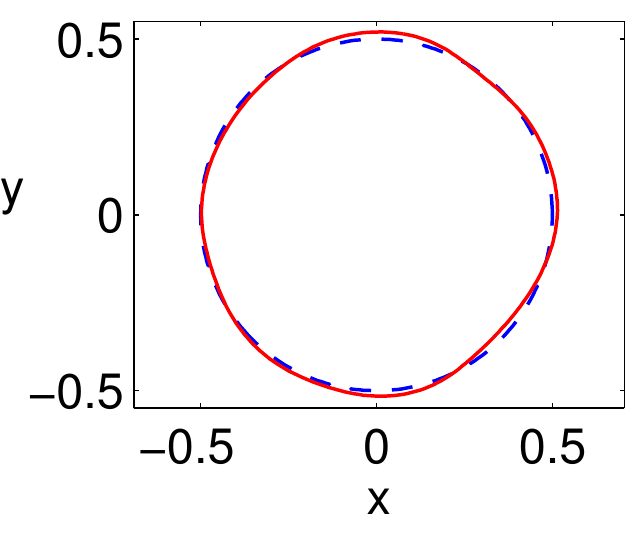}
\includegraphics[width=0.25\textwidth]{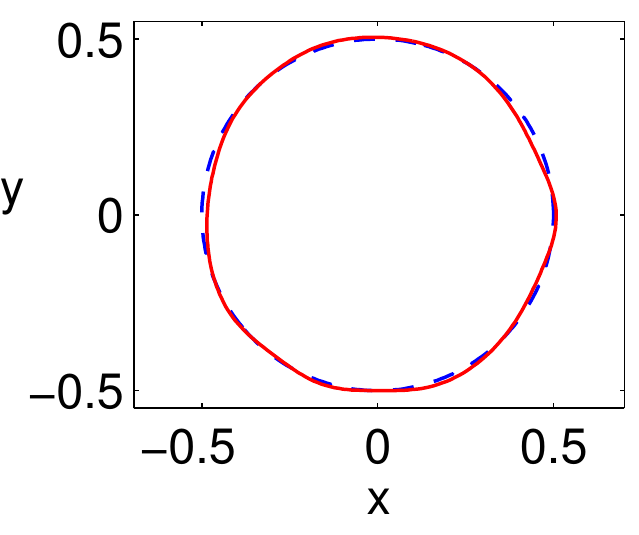}
\includegraphics[width=0.25\textwidth]{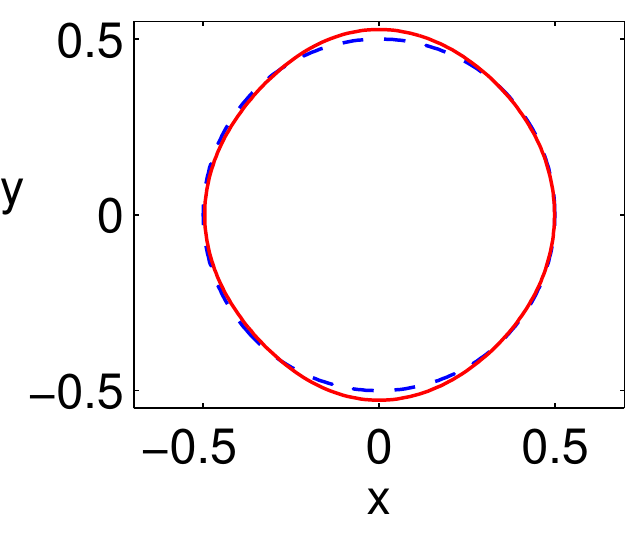}
\vspace{-0.1cm}
\caption{Wavy cylinder shapes. Left: Random waviness nr. 9, Middle: Random waviness nr. 10, Right: Optimal waviness at $Re=50$.  
Upper row: 3D illustration, with waviness amplitude exaggerated by a factor 3. Lower row: 2D wavy cross-section in scale (solid line, red online). Circular cylinder shape shown for comparison (dashed line, blue online).   
}
\label{fig:random}
\end{figure}
\begin{figure}
\vspace{-2.24cm}
\centering
\includegraphics[width=0.7\textwidth]{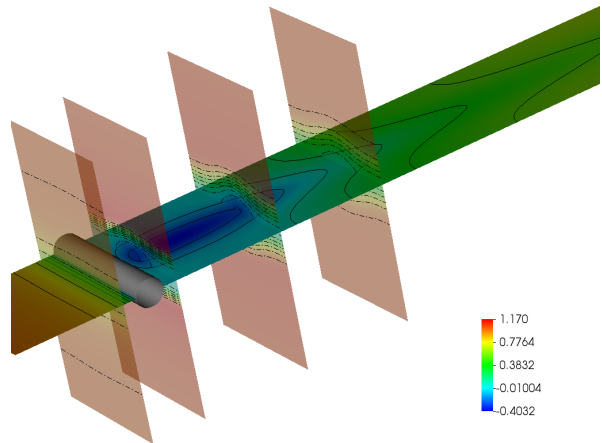} 
\vspace{-3cm}
\caption{The optimal wavy cylinder at $Re=100$ (in real scale), and streamwise velocity at selected cross-sections shown by both colours and contours with spacing $\Delta U=0.16$.}
\vspace{-0.5cm}
\label{fig:optill}
\end{figure}
\begin{figure}
\centering
\includegraphics[width=0.4\textwidth]{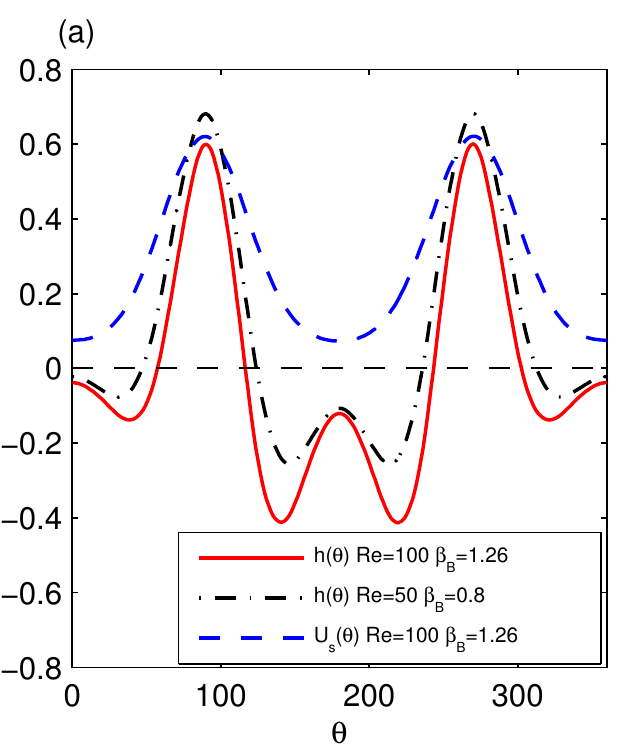}
\includegraphics[width=0.39\textwidth]{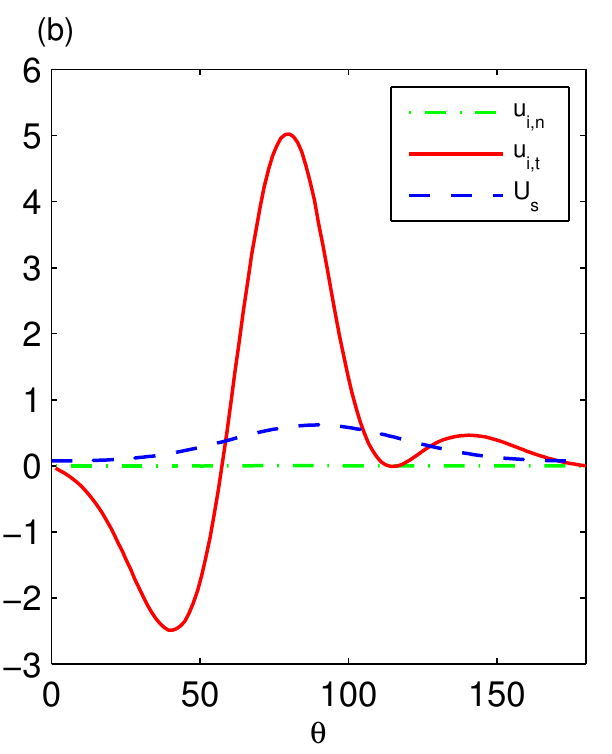} 
\caption{(a) Optimal spanwise-wavy shape distributions $h(\theta)$ compared with the optimal spanwise-wavy suction distribution $U_s$, all obtained by the method in Sec.\ \ref{sec:pert}. (b) Tangential ($u_{i,t}$) and normal ($u_{i,n}$) surface velocity distributions induced by $h(\theta)$, compared with $U_s$.}
\label{fig:indvel}
\vspace{-0.5cm}
\end{figure}

Now, we will address the optimal wavy surface shape near bifurcation at $Re=50$ for validation purposes. To make the stabilization more challenging, the wavelength of the waviness is chosen to be $\beta_B=0.8$, longer than the wavelength ($\beta_B=1-2$) which was stated to be optimal in previous studies of boundary suction. After validation, we will proceed to look at the optimal shape at $Re=100$ at the near-optimal wavelength of $\beta_B=1.26$. \footnote{It needs to be emphasized that the optimal wavelength selection results might not carry over from boundary suction to boundary shape; however, if physical stabilization mechanisms are similar one might expect qualitatively similar wavelength selection. Further wavenumber variations will be omitted here for brevity; this paper presents a method for computing a second-order optimal stabilizing distribution of waviness for a given spanwise wavenumber.}

The most stabilizing azimuthal distribution is predicted in FreeFem++ using the 2D $2^{nd}$-order sensitivity method outlined in Sec.\
 \ref{sec:pert}. Now, this optimal distribution will be validated by direct computation of eigenvalues for 3D wavy cylinders in Nek5000. The validations are presented in Table \ref{table:t1}. The straight cylinder eigenvalue at $Re=50$ in Nek5000 is found to be unstable with growth rate $0.0128$. The eigenvalue for the cylinder with predicted optimal waviness at $A=0.04$ stabilizes to $-0.0235$ in Nek5000. This shows that the predicted optimal waviness stabilizes the flow in Nek5000. 

Next, we investigate whether the predicted shape is an optimal (most stabilizing) shape, at least for small waviness amplitudes. If the predicted shape is a local optimal, then any small deviation from it should produce less stable growth rates than the optimal shape does in a full 3D (tri-global) stability computation. To confirm this, we superpose the optimal shape with a random shape distribution of a small amplitude ($A=0.01$), and re-normalise the result to $A=0.04$. This test has been done for 10 different random azimuthal shape distributions using the Gaussian basis with $N=20$, at the same spanwise wavelength $\beta_B=0.8$. The coefficient vectors $\textbf{a}$ (Eq.\ \ref{eq:decomp}) for these random shapes are the 10 first vectors of length $20$ obtained from the \textbf{rand}-function in MATLAB with the default settings. The superposition of the optimal shape with each of the random shapes was created in Nek5000, and the 3D eigenvalues computed, summarized in table \ref{table:t1}. In all cases, the eigenvalue is more stabilized by the optimal shape than the perturbed optimal. This confirms that the predicted optimal is at least a local optimal. If the theory is correct, then the predicted optimal is also a global optimal.   

It is also interesting to see how much better the optimal performs compared to random waviness with $\beta_B=0.8$. To see this, we have performed a second test where the random shape distributions were applied alone (without the optimal). The eigenvalues are again listed in table \ref{table:t1}. They show that none of the random shapes stabilizes the flow at $A=0.04$. \footnote{A few random shapes will probably stabilize the flow at higher amplitudes, as eigenvalue drift is quadratic.} The optimal decreases the growth rate three times as much as the best random wavy shape at $A=0.04$. 

The optimal distribution of waviness is illustrated in Fig.\ \ref{fig:random}, right column, alongside two of the random shapes (left and middle columns). The amplitude of the waviness is exaggerated in the 3D illustrations (top). The optimal distribution is symmetric with respect to $y=0$ ("in-phase" as expected from previous spanwise-wavy suction studies) and attains its maximum displacement near $\theta=\pm 90^{\circ}$ ($y=\pm 0.5$). 
Random shape 9 (Fig.\ \ref{fig:random} left) also has a substantial amplitude near $\theta=\pm 90^{\circ}$, and is the second most efficient with growth rate decrease $\delta \sigma_r=-0.01$.  Random shape 10 (Fig.\ \ref{fig:random} right) distributes its amplitude at the upstream and downstream ends of the cylinder, and is inefficient with $\delta \sigma_r=-0.002$. For square cylinders, both leading edge and trailing edge waviness has been considered previously \cite{darekarsherwin}. The present results for a circular cylinder indicate that waviness near the wake separation has the largest effect; if this can be generalized to a square cylinder, then the best position for the waviness would be at the corners of the trailing edge. Summarizing, the results so far show clearly two features: (i) the method produces optimal shapes to a good approximation, and (ii) the distribution of the waviness is important --- random shapes are less efficient than optimal shape at a fixed spanwise wavelength. 

Going forward, we investigate changes in the optimal shape and its performance when increasing the Reynolds number to $Re=100$. Both 2D predictions and 3D validations confirm that the waviness is even more efficient at $Re=100$ than at $Re=50$; an amplitude of A=0.02 (maximum displacement only 1$\%$ of the cylinder diameter) is sufficient to stabilize the flow in both analyses at $Re=100$. The optimal shape distributions at $Re=50$ and $Re=100$ are depicted together in Fig.\ \ref{fig:indvel}(a), showing that the optimal distribution of waviness remains qualitatively similar. 

Finally, design robustness of the optimal solution deserves to be mentioned, as if the optimality range is very narrow, the stabilizing influence might not be observed in real-life applications. Indications of the design robustness can be obtained by comparing the magnitudes of different eigenvalues of the Hessian matrix $\tilde \sigma$. The eigenvectors of the Hessian form an orthonormal basis. If the most positive (destabilizing) Hessian eigenvalue is of larger or similar magnitudes as the most negative (stabilizing) Hessian eigenvalue, then the optimal shape is not robust. The reason is that small components of destabilizing eigenvectors could counteract the stabilizing influence. At $Re=50$, the ratio between the most stabilizing and most destabilizing eigenvalues of the real part of the Hessian is $20.33$, which seems relatively robust. At $Re=100$, this ratio is $195$, which is considerably more robust. 

\subsection{Relation between the optimal shape and the optimal suction distributions}
The optimal suction distribution on a circular cylinder was considered in \cite{delguercio2014,delguercio2014b}. They observed that the suction changes the base flow by creating streaks through a lift-up effect. The suction distribution which creates streaks of the maximal amplitude for a given suction amplitude was found. The peak amplitude of this distribution occurred at $\theta=\pm90$, the same as the spanwise-wavy surface here, suggesting that the same mechanism may be active. Fig.\ \ref{fig:optill} shows the optimal wavy cylinder at $Re=100$ and $A=0.02$ in scale, together with the streamwise velocity around this cylinder from a DNS. The figure illustrates that the flow is steady, and that the wavy cylinder is a very efficient streak generator, which may explain its stabilizing influence. The surface waviness is so small that it seems unobservable. The waviness-induced streaks, \textit{i.e.} the variation of the streamwise velocity, are however strong. The strength of the streaks is shown by the spanwise variation of colours in the horizontal cross-section, and length of the streaks is indicated by the displacement of the velocity contours in the horizontal plane. The reverse flow velocity has a spanwise variation from $U=-0.4$ to $U=-0.01$ at $z \approx 3$, nearly breaking the zone into separate recirculation cells. The vertical velocity variation is small in comparison to the streamwise velocity variation, as expected for streaks. This is shown by that contours in the vertical planes are displaced very little.           

As discussed in Sec.\ \ref{sec:pert}, the effect of the shape change is modelled with a (linearized) velocity boundary condition at the original surface position: $\delta \textbf{U}_n=-h_n(\nabla \textbf{U})\cdot \hat n$ for basis function $h_n$ (see \textit{e.g.} \cite{gunzburger}). It may be interesting to compare this equivalent velocity distribution at the surface to the optimal steady suction distribution. To do this, we have computed the optimal suction distribution at $Re=100$ using the method of $2^{nd}$-order optimals. 
The obtained optimal suction distribution at $Re=100$ is shown by a dashed line in Fig.\ \ref{fig:indvel} (a,b). The distribution seems quite identical to the one shown in \cite{delguercio2014b}. While the present approach is purely mathematical, their approach was based on optimal streak generation, so this result supports their physical arguments.
The equivalent velocity distribution of the wavy cylinder surface is depicted in Fig.\ \ref{fig:indvel} (b). It is immediately clear that the tangential component (solid line, red online) dominates the normal component (dashed line, green online). 
Hence, the surface waviness acts on the tangential velocity component, by extracting energy from the base flow derivatives at the cylinder surface. Moreover, the velocity changes sign at $\theta\approx 60^\circ$, and a high tangential velocity is induced around this point. This implies that the fluid near the surface is pushed alternatively towards and away from the line $\theta\approx 60^\circ$. The suction, in contrast, changes the normal (vertical) velocity component at the top of the cylinder, which in turn influences the streamwise velocity by lift-up effect. 
%

\section{Conclusions \label{sec:concl}}
A new method to compute optimal second-order perturbations of global instability problems is formulated and applied to find the optimal spanwise-wavy cylinder surface to passively stabilize vortex shedding. The optimal distribution of waviness is found for a given spanwise wavelength. The wake around a circular cylinder is stabilized in global modes and DNS at $Re=100$ and $Re=50$, with the maximal surface displacement ca. $1\%$ and $2\%$, respectively.   
Previous results for optimal spanwise-wavy suction are also recovered but without a priori hypothesis about the physical mechanisms involved. The method is based on perturbation theory of linearized Navier-Stokes equations, and hence should be applicable for a wide class of flows where $2^{nd}$-order perturbations are relevant. 
As a next step, the analysis should be applied to other flows where the eigenvalue drift is confirmed to be of the $2^{nd}$ order, and where the boundary modulation is not due to streak generation (for example the stenotic flow in\cite{johnstenosis}). We also foresee experimental confirmation of the optimal shapes presented here.

\section*{References}

\bibliography{../../../../../Rapporter/postdocref}

\end{document}